\documentclass[showpacs,twocolumn,pre]{revtex4}
\usepackage{amsmath}
\usepackage{latexsym}
\usepackage{epsfig}
\usepackage{makeidx}
\usepackage{amsfonts}

\begin{document}

\title{Nonlinear subdiffusive fractional equations and aggregation
phenomenon }
\author{Sergei Fedotov }
\affiliation{School of Mathematics, The University of Manchester, Manchester M13 9PL, UK }

\begin{abstract}
In this article we address the problem of the nonlinear interaction of
subdiffusive particles. We introduce the random walk model in which
statistical characteristics of a random walker such as escape rate and jump
distribution depend on the mean density of particles. We derive a set of
nonlinear subdiffusive fractional master equations and consider their
diffusion approximations. We show that these equations describe the
transition from an intermediate subdiffusive regime to asymptotically normal
advection-diffusion transport regime. This transition is governed by
nonlinear tempering parameter that generalizes the standard linear
tempering. We illustrate the general results through the use of the examples
from cell and population biology. We find that a nonuniform anomalous
exponent has a strong influence on the aggregation phenomenon.
\end{abstract}

\maketitle

\section{Introduction}

Anomalous subdiffusion is a widespread phenomenon in physics and biology
\cite{MK1,Klages,MFH}. It is observed in the transport of proteins and
lipids on cell membranes \cite{Sa}, RNA molecules in the cells \cite{Go},
signaling molecules in spiny dendrites \cite{Santa}, and elsewhere. Apart
from fractional Brownian motion, the \textit{linear }fractional equations
are the standard models for the description of anomalous subdiffusive
transport \cite{Klages}. In these models the diffusing particles do not
interact. The question then arises as to how to extend these equations for
the \textit{nonlinear} case, involving particles interactions. These
nonlinear effects are typical and very important in cellular and population
biology. It is well known that many biophysical processes in microorganisms
depend on the their population density. A typical example is the \textit{%
quorum sensing phenomenon,} in which microorganisms coordinate their
behaviour according to their local population density \cite{Q}. Other
examples are a \textit{cellular adhesion,} which involves the interaction
between neighbouring cells \cite{Ad,V,Si,Ha}; and the \textit{volume filling
effect,} which describes the dependence of cell motility on the availability
of space in a crowded environment \cite{Hillen,Hillen2}. The understanding
of macroscopic phenomena like cell and microorganism aggregation requires an
understanding of how individual species interact through attractive or
repulsive forces (see, for example, \cite{Campos} and references therein).
The attraction between individuals may result from various social
interactions such as mating, settlement, defense against predators, etc.
While the repulsion may occur due to low resources in highly populated
regions \cite{Campos,Fedot5}. Note that microorganisms interact both
directly and indirectly via signaling molecules.

The main purpose of this paper is to incorporate these nonlinear effects
into subdiffusive equations. Our aim is to take into account the interaction
between particles on the mesoscopic level, at which the random walker's
characteristics depend on the mean field density of particles. Our intention
is to derive the subdiffusive nonlinear fractional equations for the density
of particles and apply these equations to the problem of aggregation. In
this paper we use two different approaches that are based on the
density-dependent dispersal kernels and density-dependent jump rate. Note
that several theoretical studies have been devoted to nonlinear
generalizations of linear fractional equations. However, most research has
been focused on the problem how to incorporate the nonlinear reactions into
subdiffusive equations \cite%
{VR,YH,Nec,Sokolov,Fedot1,Abad,Fedot3,Soko,Sh,Vo,Au}. The aim of this paper
is to study the nonlinear subdiffusive transport processes involving the
anomalous trapping of particles and their interactions.

In this paper we deal with the random walk model involving a residence time
dependent escape rate and the structural density of particles. This has been
used by many authors for the analysis of the non-Markovian random walks \cite%
{MFH,VR,YH,Fedotov,Steve1,Zubarev}. It turns out that this linear model is
the most suitable for further nonlinear generalizations. We consider a
`space-jump' random walk in one space dimension. The particle waits for a
random time (residence time) $T_{x}$ at point $x$ in space before making a
jump to another point. The random residence time $T_{x}$ is determined by
the probability density function $\psi (x,\tau )=\Pr \left\{ \tau
<T_{x}<\tau +d\tau \right\} .$ The key characteristic of this random walk is
the escape rate $\gamma $ from the point $x.$ It depends on the residence
time $\tau $ and the position $x:$ $\gamma =\gamma (x,\tau ).$ This rate can
be rewritten in terms of the probability density function $\psi (x,\tau )$
and the survival probability $\Psi (x,\tau )=\int_{\tau }^{\infty }\psi
(x,u)du$ as follows \cite{Cox}%
\begin{equation}
\gamma (x,\tau )=\frac{\psi (x,\tau )}{\Psi (x,\tau )}.  \label{gamma}
\end{equation}%
It is convenient to write the survival probability $\Psi (x,\tau )$ in terms
of $\gamma (x,\tau )$ as follows
\begin{equation}
\Psi (x,\tau )=e^{-\int_{0}^{\tau }\gamma (x,s)ds}.  \label{sur}
\end{equation}

Let $\xi (x,\tau ,t)$ be the structural density of particles at point $x$ at
time $t$ whose residence time lies in the interval $\left( \tau ,\tau +d\tau
\right) $. Here we neglect the aging phenomenon \cite{Bar2} and assume that
at $t=0$ all particles have zero residence time:
\begin{equation}
\xi (x,\tau ,0)=\rho _{0}(x)\delta (\tau ),  \label{zero}
\end{equation}%
where $\rho _{0}(x)$ is the initial density. To consider the aging we need
to specify the general distribution for $\xi (x,\tau ,0).$

The density of particles, $\rho \left( x,t\right) ,$ is obtained by
integration of the structural density $\xi (x,\tau ,t)$ with respect to the
residence time variable $\tau $ from $0$ to $t$:%
\begin{equation}
\rho \left( x,t\right) =\int_{0}^{t}\xi (x,\tau ,t)d\tau .  \label{den}
\end{equation}%
The number of particles with fixed residence time $\tau $ escaping from the
point $x$ per unit of time is defined as a product $\gamma (x,\tau )\xi
(x,\tau ,t).$ The total escape rate, $i\left( x,t\right) ,$ of particles
from the point $x$ can be obtained by integration of this product with
respect to $\tau $ from $0$ to $t$:
\begin{equation}
i(x,t)=\int_{0}^{t}\gamma (x,\tau )\xi (x,\tau ,t)d\tau .  \label{mean}
\end{equation}%
The rate $i\left( x,t\right) $ is very useful quantity, since it allows to
write a very simple master equation for the density $\rho \left( x,t\right) $
as the balance of particles at the point $x$
\begin{equation}
\frac{\partial \rho }{\partial t}=\int_{\mathbb{R}}i\left( x-z,t\right)
w\left( z|x-z\right) dz-i\left( x,t\right) ,  \label{mast}
\end{equation}%
where $w\left( z|x\right) $ is the dispersal kernel for the jumps. We have
assumed here that the jumps of particles are independent from the residence
time. One of the main results in the non-Markovian random walk theory is
that the mean escape rate $i\left( x,t\right) $ can be written as a
convolution
\begin{equation}
i\left( x,t\right) =\int_{0}^{t}K\left( x,t-\tau \right) \rho \left( x,\tau
\right) d\tau .  \label{m_escape}
\end{equation}%
(see, for example, \cite{MFH}). Here $K(x,t)$ is the memory kernel defined
by its Laplace transform
\begin{equation}
\hat{K}\left( x,s\right) =\frac{\hat{\psi}\left( x,s\right) }{\hat{\Psi}%
\left( x,s\right) },  \label{L}
\end{equation}%
where $\hat{\psi}\left( x,s\right) =\int_{0}^{\infty }e^{-s\tau }\psi
(x,\tau )d\tau $ and $\hat{\Psi}\left( x,s\right) =\int_{0}^{\infty
}e^{-s\tau }\Psi (x,\tau )d\tau $.

In the anomalous subdiffusive case, the survival probability $\Psi _{\mu
}\left( x,\tau \right) $ can be modelled by the Mittag-Leffler function \cite%
{Scalas}
\begin{equation}
\Psi _{\mu }\left( x,\tau \right) =E_{\mu }\left[ -\left( \frac{\tau }{\tau
_{0}(x)}\right) ^{\mu (x)}\right] ,\ 0<\mu (x)<1,  \label{sub2}
\end{equation}%
where $\mu (x)$ is the space-dependent anomalous exponent. In what follows
we assume for simplicity that $\tau _{0}$ is constant. The Laplace transform
of the memory kernel $K_{\mu }\left( x,t\right) $ is $\hat{K}_{\mu }\left(
x,s\right) =s^{1-\mu (x)}\tau _{0}{}^{-\mu (x)}$ and the integral anomalous
escape rate $i\left( x,t\right) $ can be written as
\begin{equation}
i\left( x,t\right) =\frac{1}{\tau _{0}{}^{\mu (x)}}\mathcal{D}_{t}^{1-\mu
(x)}\rho \left( x,t\right) ,  \label{ano_escape}
\end{equation}%
where $\mathcal{D}_{t}^{1-\mu (x)}$ is the Riemann-Liouville derivative with
varying anomalous exponent \cite{Ch}. Substitution of (\ref{ano_escape})
into (\ref{mast}) gives the integral fractional equation with
space-dependent anomalous exponent
\begin{eqnarray}
\frac{\partial \rho }{\partial t} &=&\int_{\mathbb{R}}\frac{1}{\tau
_{0}{}^{\mu (x-z)}}\mathcal{D}_{t}^{1-\mu (x-z)}\rho \left( x-z,t\right)
w\left( z|x-z\right) dz  \notag \\
&&-\frac{1}{\tau _{0}{}^{\mu (x)}}\mathcal{D}_{t}^{1-\mu (x)}\rho \left(
x,t\right) .  \label{mm}
\end{eqnarray}%
Note that if $\mu =const$ and $\tau _{0}=1$ one obtains the following
equation \cite{Ma}
\begin{equation}
\frac{\partial ^{\mu }\rho }{\partial t^{\mu }}=\int_{\mathbb{R}}\rho \left(
x-z,t\right) w\left( z|x-z\right) dz-\rho \left( x,t\right) ,  \label{Caputo}
\end{equation}%
where the Caputo derivative $\partial ^{\mu }\rho /$ $\partial t^{\mu }$ is
used instead of the Riemann-Liouville derivative $\mathcal{D}_{t}^{1-\mu
}\rho $.

Using the Taylor series expansion in terms of $z,$ and a symmetric dispersal
kernel $w$ for which $\int_{\mathbb{R}}zw(z|x)dz=0$, we obtain the standard
fractional subdiffusive equation%
\begin{equation}
\frac{\partial \rho }{\partial t}=\frac{\partial ^{2}}{\partial x^{2}}\left[
D_{\mu }\left( x\right) \mathcal{D}_{t}^{1-\mu (x)}\rho \left( x,t\right) %
\right]  \label{fr}
\end{equation}%
with the fractional diffusion coefficient
\begin{equation*}
D_{\mu }\left( x\right) \mathcal{=}\frac{\sigma ^{2}(x)}{2\tau _{0}{}^{\mu
(x)}},
\end{equation*}%
where $\sigma ^{2}(x)=\int_{\mathbb{R}}z^{2}w(z|x)dz$.

It has been found recently that subdiffusive fractional equations with
constant $\mu $ are not structurally stable with respect to the spatial
variations of fractal exponent $\mu (x)$ \cite{Steve1,Zubarev}. This leads
to the anomalous aggregation of particles at the minimum of the function $%
\mu (x)$. In heterogeneous biological systems, in which the exponent $\mu
(x) $ is space-dependent, the question arises as to whether this anomalous
aggregation of a population can be prevented. Therefore it is an important
problem to find the way how to regularize subdiffusive fractional equations.
One way is to incorporate random killing, which ensures regular behaviour in
the long-time limit \cite{Steve2}. The aim of this paper is to address this
problem through a nonlinear escape rate that takes into account repulsive
forces between particles. In the next section we derive the nonlinear
generalization of the fractional equations like (\ref{mm}) and (\ref{fr}).

\section{Non-Markovian and subdiffusive nonlinear fractional equations}

There are two major ways in which nonlinear density-dependence effects can
be implemented into non-Markovian and subdiffusive transport equations. The
simplest way is to take into account the dependence of jump density $w$ on $%
\rho $. This dependence can take into account various nonlinear effects such
as adhesion, quorum sensing, volume filling, etc. However, we begin with
more complicated case of the random walk model for which the escape rate is
a function of the residence time and the local density of particles.

\subsection{Nonlinear escape rate}

The main problem with this anomalous escape rate (\ref{ano_escape}) is the
phenomenon of anomalous aggregation \cite{Fedotov,Steve1,Zubarev}.
Nonuniform distribution of the anomalous exponent $\mu (x)$ over the finite
domain $\left[ 0,L\right] $ leads to%
\begin{equation}
\rho \left( x,t\right) \rightarrow \delta (x-x_{M})\qquad as\qquad
t\rightarrow \infty .  \label{co}
\end{equation}%
Here $x_{M}$ is the point in space where the anomalous exponent $\mu \left(
x\right) $ has a minimum. The problem is that that the escape rate (\ref%
{ano_escape}) is a linear functional of the density of particles $\rho
\left( x,t\right) $ and does not take into account nonlinear effects of
repulsive forces which, in many situations, can prevent anomalous
aggregation. According to (\ref{co}) all particles aggregate into a small
region around the point $x=x_{M}$ forming a high density system. To prevent
such anomalous aggregation, one can assume that the overcrowding leads to an
increase of repulsive forces and a corresponding correction of the anomalous
escape rate $\gamma (x,\tau ).$

We assume that the probability of escape due to the repulsive forces is
independent from anomalous trapping. We define this probability for a small
time interval $\Delta t$ as
\begin{equation}
\alpha (\rho (x,t))\Delta t+o(\Delta t).
\end{equation}%
Here $\alpha (\rho )$ is the transition rate which is an increasing function
of the particles density $\rho $. Another interpretation can be given in
terms of the quorum sensing phenomenon \cite{Q}. The large cell density can
lead to the local over-depletion of nutrients and oxygen and as a result
cells can change phenotype from a proliferating state to a migrating one
\cite{Fedot3}. In another words, when the concentration of cells is low, $%
\alpha (\rho )=0$, \ but if the concentration of cells $\rho (x,t)$ reaches
a certain level $\rho _{cr}:$ $\alpha (\rho )\neq 0$ for $\rho \geq \rho
_{cr}.$ Of course, one can assume a non-monotonic dependence of the
transition rate $\alpha (\rho )$ on $\rho $. For example, at low cell
densities $\alpha (\rho )$ could decrease with $\rho $, while at high
densities $\alpha (\rho )$ could be an increasing function: $\alpha (\rho
)=\alpha _{0}\left( 1-a_{1}\rho +a_{2}\rho ^{2}\right) $ \cite{Campos,Hillen}%
.

Firstly we formulate the Markovian model for the random walk of particles
with the density-dependent escape rate. The balance equation for the
structural density $\xi (x,\tau ,t)$ takes the form
\begin{eqnarray*}
\xi (x,\tau +\Delta t,t+\Delta t,) &=&\xi (x,\tau ,t)\left( 1-\gamma (x,\tau
)\Delta t\right) \\
&&\times \left( 1-\alpha (\rho (x,t)\right) \Delta t)+o\left( \Delta
t\right) .
\end{eqnarray*}%
where $1-\gamma (x,\tau )\Delta t$ is the survival probability during time $%
\Delta t$ due to the trapping, and $1-\alpha (\rho (x,t))\Delta t$ is the
survival probability during time $\Delta t$ corresponding to the repulsion
forces between particles. In the limit $\Delta t\rightarrow 0,$ we obtain
\begin{equation}
\frac{\partial \xi }{\partial t}+\frac{\partial \xi }{\partial \tau }=-\left[
\gamma (x,\tau )+\alpha (\rho (x,t))\right] \xi ,  \label{dif1}
\end{equation}%
where the effective transition rate is the sum of two escape rates:
\begin{equation}
\gamma (x,\tau )+\alpha (\rho (x,t)).  \label{two}
\end{equation}%
The second term $\alpha (\rho (x,t))$ can be treated as the correction of
the escape rate $\gamma (x,\tau )$ defined by (\ref{gamma}) due to repulsive
forces. The boundary condition for $\xi (x,\tau ,t)$ at zero residence time $%
\tau =0$ is
\begin{equation}
\xi (x,0,t)=\int_{\mathbb{R}}i(x-z,t)w\left( z|\rho (x-z,t)\right) dz.
\label{bou3}
\end{equation}%
This equation describes the balance of particles just arriving at the point $%
x$ from the different positions $x-z.$ The jumps of particles are determined
by the conditional probability density function (dispersal kernel) $w\left(
z|\rho (x-z,t)\right) $ that depends on the total density of particles $\rho
$ at the point $x-z$ from which the particles jump at point $x$ (see all
details regarding $w$ in subsection D). Obviously this dispersal kernel
satisfies the normalization condition
\begin{equation*}
\int_{\mathbb{R}}w\left( z|\rho (x,t)\right) dz=1.
\end{equation*}%
Because of (\ref{two}) and (\ref{den}), the effective escape rate $i\left(
x,t\right) $ from the point $x$ at time $t$ can be written as%
\begin{equation}
i\left( x,t\right) =\int_{0}^{t}\gamma (x,\tau )\xi (x,\tau ,t)d\tau +\alpha
(\rho )\rho (x,t).  \label{eff}
\end{equation}%
It is convenient to introduce the number of particles $j(x,t)$ just jumping
at the point $x$ at time $t:j(x,t)=\xi (x,0,t)$, so the boundary condition (%
\ref{bou3}) can be rewritten as
\begin{eqnarray}
j(x,t) &=&\int_{\mathbb{R}}\int_{0}^{t}\gamma (x-z,\tau )\xi (x-z,\tau ,t)
\notag \\
&&\times w\left( z|\rho (x-z,t)\right) d\tau dz  \notag \\
&&+\int_{\mathbb{R}}\alpha (\rho (x-z,t))\rho (x-z,t)  \notag \\
&&\times w\left( z|\rho (x-z,t)\right) dz.  \label{bound2}
\end{eqnarray}%
We solve (\ref{dif1}) for $\xi (x,\tau ,t)$ by the methods of
characteristics. For $\tau <t$ we find
\begin{equation}
\xi (x,\tau ,t)=\xi (x,0,t-\tau )e^{-\int_{0}^{\tau }\gamma (x,\tau
)ds-\int_{t-\tau }^{t}\alpha (\rho (x,s))ds}.  \label{sol3}
\end{equation}%
This solution involves an exponential factor $e^{-\int_{0}^{\tau }\gamma
(x,\tau )ds}$ that can be interpreted as the survival function $\Psi (x,\tau
)$ defined by (\ref{sur}). Taking into account the initial condition (\ref%
{zero}), (\ref{gamma}) and substituting (\ref{sol3}) into (\ref{bound2}), we
obtain
\begin{eqnarray}
j(x,t) &=&\int_{\mathbb{R}}\int_{0}^{t}[\psi (x-z,\tau )j(x-z,t-\tau )
\label{jjjj} \\
&&\times e^{-\int_{t-\tau }^{t}\alpha (\rho (x,s))ds}w\left( z|\rho
(x-z,t)\right) d\tau dz  \notag \\
&&+\int_{\mathbb{R}}[\alpha (\rho (x-z,t))\rho (x-z,t)+\rho _{0}(x-z)  \notag
\\
&&\times \psi (x-z,t)e^{-\int_{0}^{t}\alpha (\rho (x,s))ds}]w\left( z|\rho
(x-z,t)\right) dz.  \notag
\end{eqnarray}%
Substitution of (\ref{sol3}) and (\ref{zero}) into (\ref{den}) gives%
\begin{eqnarray}
\rho (x,t) &=&\int_{0}^{t}[\Psi (x,\tau )j(x,t-\tau )e^{-\int_{t-\tau
}^{t}\alpha (\rho (x,s))ds}d\tau  \notag \\
&&+\Psi (x,t)\rho _{0}(x)e^{-\int_{0}^{t}\alpha (\rho (x,s))ds}.  \label{rr}
\end{eqnarray}%
By using the Laplace transforms one can eliminate $j(x,t)$ from the above
equations and express the the \textit{integral} escape rate $i\left(
x,t\right) $ in terms of the density $\rho (x,t)$ as
\begin{eqnarray}
i\left( x,t\right) &=&\int_{0}^{t}K\left( x,t-\tau \right) e^{-\int_{\tau
}^{t}\alpha (\rho (x,s))ds}\rho \left( x,\tau \right) d\tau  \notag \\
&&+\alpha (\rho (x,t))\rho (x,t),  \label{i2}
\end{eqnarray}%
where the memory kernel $K\left( x,t\right) $ is defined by its Laplace
transform (\ref{L}). The details of this derivation can be found in \cite%
{Fedot1,MFH}. Note that the first term in (\ref{i2}) involves the
exponential factor, $e^{-\int_{\tau }^{t}\alpha (\rho (x,s))ds},$ which can
be interpreted as a nonlinear tempering. The main feature of this modified
escape rates is that although the \textit{local} rates $\gamma (x,\tau )$\
and $\alpha (\rho (x,t))$\ are additive (see (\ref{two})), the corresponding
terms in the \textit{integral} escape rate (\ref{i2}) are not additive. This
is clearly non-Markovian effect. One of the main aims of this paper is to
find out what the implications of this effect on the long-time behaviour of
the density $\rho (x,t)$ are (see section III). Note that in the linear
homogeneous case, when $\alpha (\rho )=0,$ $\psi $ and $w$ are independent
of $x$, we obtain from Eqs. (\ref{jjjj}), (\ref{rr}) and (\ref{i2}) the
classical continuos time random walk (CTRW) equation \cite{MK1,Klages,MFH}%
\begin{equation*}
\rho (x,t)=\int_{\mathbb{R}}\int_{0}^{t}\rho (x-z,t-\tau )\psi (\tau
)w(z)d\tau dz+\rho _{0}(x)\Psi (t).
\end{equation*}

\subsection{ Master equation for the mean field density of particles}

The master equation for the density $\rho \left( x,t\right) $ takes the
simple form of the balance of jumping particles
\begin{equation}
\frac{\partial \rho }{\partial t}=\int_{\mathbb{R}}i\left( x-z,t\right)
w\left( z|\rho (x-z,t)\right) dz-i\left( x,t\right) ,  \label{Master444}
\end{equation}%
where the dispersal kernel $w$ depends on the mean density $\rho (x,t).$
Substitution of the integral escape rate (\ref{i2}) into the equation (\ref%
{Master444}) gives the closed equation for $\rho $%
\begin{eqnarray}
\frac{\partial \rho }{\partial t} &=&\int_{\mathbb{R}}\int_{0}^{t}K\left(
x-z,t-\tau \right) e^{-\int_{\tau }^{t}\alpha (\rho (x-z,s))ds}  \notag \\
&&\times \rho \left( x-z,\tau \right) w\left( z|\rho (x-z,t)\right) d\tau dz
\notag \\
&&-\int_{0}^{t}K\left( x,t-\tau \right) e^{-\int_{\tau }^{t}\alpha (\rho
(x,s))ds}\rho \left( x,\tau \right) d\tau  \notag \\
&&+\int_{\mathbb{R}}\alpha (\rho (x-z,t))\rho (x-z,t)w\left( z|\rho
(x-z,t)\right) dz  \notag \\
&&-\alpha (\rho (x,t))\rho (x,t).  \label{Mastt}
\end{eqnarray}%
One can find various nonlinear diffusion approximations of (\ref{Mastt})
assuming the particular expressions for the density-dependent dispersal
kernels $w\left( z|\rho \left( x,t\right) \right) $ and density-dependent
jump rate $\alpha (\rho \left( x,t\right) )$ \cite{Campos}.

When the escape rate $\gamma $\ does not depend on the residence time
variable $\tau $, the survival function $\Psi (x,\tau )$ (\ref{sur}) has an
exponential form: $\Psi (x,\tau )=e^{-\gamma (x)\tau }$. In this Markovian
case the Laplace transform of the memory kernel, $\hat{K}\left( x,s\right) ,$
does not depend on $s:$ $\hat{K}\left( x,s\right) =\gamma (x).$ The integral
escape rate $i(x,t)$ takes the standard Markovian form:%
\begin{equation}
i(x,t)=\left[ \gamma (x)+\alpha (\rho (x,t))\right] \rho \left( x,t\right) .
\end{equation}%
Substitution of this formula into (\ref{Master444}) gives the nonlinear
Kolmogorov-Feller equation for $\rho \left( x,t\right) $ \cite{MFH}%
\begin{eqnarray*}
\frac{\partial \rho }{\partial t} &=&\int_{\mathbb{R}}[\gamma (x-z)+\alpha
(\rho (x-z,t))]\rho \left( x-z,\tau \right) \\
&&\times w\left( z|\rho (x-z,t)\right) dz-\gamma (x)\rho -\alpha (\rho )\rho
.
\end{eqnarray*}%
Several approximations of this equation and its applications in population
biology have been discussed in \cite{Campos}.

\subsection{Diffusion approximation and chemotaxis}

In this subsection we derive from (\ref{Mastt}) a fractional subdiffusive
equation for $\rho $ when $\alpha (\rho )=0$. The main motivation is to
study the chemotaxis which is a directed migration of cells toward a more
favorable environment \cite{Hillen2,OS}.The aim is to illustrate as to how a
fractional chemotaxis equation for cell movement can be derived. We consider
the random walk in which particle (cell) performing instantaneous jumps in
space such that the jump density $w$ involves only two outcomes
\begin{equation}
w\left( z|x,t\right) =r(x,t)\delta (z-a)+l(x,t)\delta (z+a),  \label{ww}
\end{equation}%
where $a$ is the jump size, $r(x,t)$ is the jump probability from the point $%
x$ to $x+a$, $l(x,t)$ is the jump probability from the point $x$ to $x-a$
and
\begin{equation}
r(x,t)+l(x,t)=1.
\end{equation}%
For the jump kernel (\ref{ww}) the master equation (\ref{Master444}) takes
the form
\begin{eqnarray}
\frac{\partial \rho }{\partial t} &=&r(x-a,t)i(x-a,t)  \notag \\
&&+l(x+a,t)i(x+a,t)-i(x,t).
\end{eqnarray}%
In the limit $a\rightarrow 0,$ we obtain
\begin{eqnarray}
\frac{\partial \rho }{\partial t} &=&-a\frac{\partial }{\partial x}\left\{
[r(x,t)-l(x,t)]i(x,t)\right\}  \notag \\
&&+\frac{a^{2}}{2}\frac{\partial ^{2}i(x,t)}{\partial x^{2}}+o(a^{2}).
\label{TTT}
\end{eqnarray}%
One can introduce the density of chemotactic substance $U(x,t)$ that induces
the movement of the particles (cells) up or down the gradient \cite{Hillen2}%
. The presence of non-zero gradient $\partial U/\partial x$ gives rise to
the bias of the random walk when $r(x,t)\neq l(x,t)$ \cite{OS,LH}$.$ We
define the difference $r(x,t)-l(x,t)$ as \cite{HLS}
\begin{equation}
r(x,t)-l(x,t)=-\beta a\frac{\partial U(x,t)}{\partial x}+o(a),  \label{www}
\end{equation}%
where $\beta $ is the measure of the strength of chemotactic movement. When $%
\beta $ is negative, the advection (taxis) is in the direction of increase
in the chemotactic substance $U(x,t)$. The equation (\ref{TTT}) can be
rewritten in terms of $U(x,t)$ as follows
\begin{equation}
\frac{\partial \rho }{\partial t}=a^{2}\frac{\partial }{\partial x}\left[
\beta \frac{\partial U}{\partial x}i(x,t)\right] +\frac{a^{2}}{2}\frac{%
\partial ^{2}i(x,t)}{\partial x^{2}}+o(a^{2}).  \label{ii}
\end{equation}%
Various expressions for the integral escape rate $i(x,t)$ generate the set
of the equations for $\rho $ in the diffusion approximations. For example,
the Markovian total escape rate
\begin{equation*}
i(x,t)=\gamma \rho (x,t)
\end{equation*}%
with escape rate $\gamma \rightarrow \infty $ and jump size $a\rightarrow 0$
give the standard advection-diffusion equation or classical Fokker-Planck
equation
\begin{equation*}
\frac{\partial \rho }{\partial t}=2D\frac{\partial }{\partial x}\left[ \beta
\frac{\partial U}{\partial x}\rho \right] +D\frac{\partial ^{2}\rho }{%
\partial x^{2}}
\end{equation*}%
with finite diffusion coefficient $D=a^{2}\gamma /2.$ Note that if we
interpret $U(x,t)$ as the external potential then $\beta ^{-1}=2kT$ \cite%
{Bar}. The anomalous escape rate (\ref{ano_escape})
\begin{equation*}
i\left( x,t\right) =\frac{1}{\tau _{0}{}^{\mu (x)}}\mathcal{D}_{t}^{1-\mu
(x)}\rho \left( x,t\right)
\end{equation*}%
generates the subdiffusive advection-diffusion equation or the fractional
Fokker-Planck equation \cite{Bar,HLS}
\begin{align}
\frac{\partial \rho }{\partial t}& =2\frac{\partial }{\partial x}\left[
D_{\mu }(x)\beta \frac{\partial U}{\partial x}\mathcal{D}_{t}^{1-\mu
(x)}\rho \right]  \notag \\
& +\frac{\partial ^{2}}{\partial x^{2}}\left[ D_{\mu }(x)\mathcal{D}%
_{t}^{1-\mu (x)}\rho \right] .  \label{ff}
\end{align}%
Here $D_{\mu }(x)$ is the fractional diffusion coefficient defined as
\begin{equation}
D_{\mu }(x)=\frac{a^{2}}{2\tau _{0}{}^{\mu (x)}}  \label{DD}
\end{equation}%
which is finite in the limit $a\rightarrow 0$ and $\tau _{0}\rightarrow 0.$

\subsection{Nonlinear jump distributions}

In this subsection we discuss various nonlinear jump distributions leading
to nonlinear fractional equations when $\alpha (\rho )=0$. As we mentioned
previously, this is the simplest way to incorporate nonlinearity into
subdiffusive fractional equations \cite{Steve1}. As long as the escape rate $%
i\left( x,t\right) $ is determined, we can define \textquotedblleft where to
jump" through the dispersal kernel $w$. Let us assume that the dispersal
kernel $w$ depends on the density of particles: $w=w(z|\rho (x,t))$. In what
follows we restrict ourselves to the subdiffusive case for which the rate $%
i(x,t)$ is determined by (\ref{ano_escape}). In this case, the starting
master equation is
\begin{eqnarray}
\frac{\partial \rho }{\partial t} &=&\int_{\mathbb{R}}\frac{\mathcal{D}%
_{t}^{1-\mu (x-z)}\rho \left( x-z,t\right) }{\tau _{0}{}^{\mu (x-z)}}w\left(
z|\rho (x-z,t)\right) dz  \notag \\
&&-\frac{\mathcal{D}_{t}^{1-\mu (x)}\rho \left( x,t\right) }{\tau
_{0}{}^{\mu (x)}}.  \label{mmm}
\end{eqnarray}

First we consider the case for which the jump dependence on $\rho $ is
local. For example, one can use a Gaussian dispersal kernel with rapidly
decaying tails as
\begin{equation*}
w\left( z|\rho \right) =\frac{1}{\sqrt{2\pi \sigma ^{2}(\rho )}}\exp \left[ -%
\frac{z^{2}}{2\sigma ^{2}(\rho )}\right] .
\end{equation*}%
An increasing dispersion $\sigma ^{2}(\rho )$ describes the effect of the
local repulsive forces due to overcrowding, while a decaying function $%
\sigma ^{2}(\rho )$ corresponds to attractive forces. Using the Taylor
series expansion in terms of $z$ in the master equation (\ref{mmm}), we
obtain the subdiffusive equation%
\begin{equation}
\frac{\partial \rho }{\partial t}=\frac{\partial ^{2}}{\partial x^{2}}\left[
K_{\mu }\left( \rho \right) \mathcal{D}_{t}^{1-\mu (x)}\rho \left(
x,t\right) \right]
\end{equation}%
with the nonlinear fractional diffusion coefficient
\begin{equation*}
K_{\mu }\left( \rho \right) =\frac{\sigma ^{2}(\rho )}{\tau _{0}{}^{\mu (x)}}%
.
\end{equation*}%
One can also introduce the nonlinear drift term generated by Bernoulli jump
distribution \cite{Campos}
\begin{eqnarray*}
w\left( z|\rho \right) &=&\frac{1}{2}\left( 1+au\left( \rho \right) \right)
\delta \left( z-a\right) \\
&&+\frac{1}{2}\left( 1-au\left( \rho \right) \right) \delta \left(
z+a\right) ,
\end{eqnarray*}%
where the positive function $u\left( \rho \right) $ takes into account the
fact that the jump on the right is more likely than a jump to the left; $a$
is the jump size. It is assumed that $1-au\left( \rho \right) $ is not
negative; that is, $u\left( \rho \right) $ $\leq a^{-1}$. The fractional
master equation (\ref{mmm}) takes the form
\begin{eqnarray}
\frac{\partial \rho }{\partial t} &=&\frac{1+au\left( \rho \left(
x-a,t\right) \right) }{2\tau _{0}{}^{\mu (x-a)}}\mathcal{D}_{t}^{1-\mu
(x-a)}\rho \left( x-a,t\right)  \notag \\
&&+\frac{1-au\left( \rho \left( x+a,t\right) \right) }{2\tau _{0}{}^{\mu
(x+a)}}\mathcal{D}_{t}^{1-\mu (x+a)}\rho \left( x+a,t\right)  \notag \\
&&-\frac{1}{\tau _{0}{}^{\mu (x)}}\mathcal{D}_{t}^{1-\mu (x)}\rho \left(
x,t\right) .  \label{bbb}
\end{eqnarray}%
Using the Taylor series expansion in (\ref{bbb}) as $a\rightarrow 0$, we
obtain the nonlinear fractional Fokker-Planck equation
\begin{align}
\frac{\partial \rho }{\partial t}& =-\frac{\partial }{\partial x}\left[
2D_{\mu }(x)u\left( \rho \left( x,t\right) \right) \mathcal{D}_{t}^{1-\mu
(x)}\rho \right]  \notag \\
& +\frac{\partial ^{2}}{\partial x^{2}}\left[ D_{\mu }(x)\mathcal{D}%
_{t}^{1-\mu (x)}\rho \right]
\end{align}%
with the \textit{nonlinear} advection term involving $u\left( \rho \left(
x,t\right) \right) $ and fractional diffusion coefficient $D_{\mu }(x).$

Now let us consider the random jump model when the jump kernel $w\left(
z|\rho \right) $ depends on the mean density at a neighbouring points
(nonlocal nonlinearity). This model deals with so-called volume filling
effect and cell-to-cell adhesion \cite{V,Hillen}. Instead of considering the
jump rates as in \cite{V,Hillen} we model volume filling and adhesion
effects via the jump density
\begin{equation*}
w\left( z|\rho \right) =r(\rho )\delta \left( z-a\right) +l(\rho )\delta
\left( z+a\right) ,
\end{equation*}%
where $r(\rho (x,t))$ is the probability of jumping right from $x$ to $x+a$
at time $t$ and $l(\rho (x,t))$ is the probability of jumping left from $x$
to $x-a$ at time $t.$ We define these probabilities by using two decreasing
functions $f_{v}(\rho )\geq 0$ and $f_{a}(\rho )\geq 0$ as follows

\begin{eqnarray}
r(\rho (x,t)) &=&\frac{f_{v}(\rho (x+a,t))f_{a}(\rho (x-a,t))}{F\left( \rho
\right) },  \notag \\
l(\rho (x,t)) &=&\frac{f_{v}(\rho (x-a,t))f_{a}(\rho (x+a,t))}{F\left( \rho
\right) },  \label{pro}
\end{eqnarray}%
where the function
\begin{eqnarray*}
F\left( \rho \right) &=&f_{v}(\rho (x-a,t))f_{a}(\rho (x+a,t)) \\
&&+f_{v}(\rho (x+a,t))f_{a}(\rho (x-a,t))
\end{eqnarray*}%
makes sure that
\begin{equation*}
r(\rho (x,t))+l(\rho (x,t))=1.
\end{equation*}%
It follows from (\ref{pro}) that the probabilities $r(\rho )$ and $l(\rho )$
of jumping into a neighbouring points are dependent on the densities of
particles at these points. The decreasing function $f_{v}(\rho (x+a,t))$ is
used to model volume filling: the jump probability $r(\rho (x,t))$ from the
point $x$ to the right $x+a$ is reduced by the the presence of cells at the
point $x+a$. The decreasing function\ $f_{a}(\rho (x-a,t))$ describes the
adhesion effect, which says that the jump probability $r(\rho (x,t))$ from
the point $x$ to the right $x+a$ is reduced by the presence of cells at the
point $x-a.$

In the limit $a\rightarrow 0$, one can obtain the nonlinear fractional
diffusion equation. To illustrate our theory let us consider the particular
case involving only volume filling effects (no adhesion):%
\begin{equation}
f_{v}(\rho )=1-\rho ,\quad f_{a}(\rho )=1.  \label{fff}
\end{equation}%
In this case, we require that the initial density
\begin{equation*}
\rho _{0}(x)<1.
\end{equation*}%
Substitution of (\ref{fff}) into (\ref{pro}) gives the following
probabilities
\begin{eqnarray}
r(\rho (x,t)) &=&\frac{1-\rho (x+a,t)}{2-\rho (x+a,t)-\rho (x-a,t)},  \notag
\\
l(\rho (x,t)) &=&\frac{1-\rho (x-a,t)}{2-\rho (x+a,t)-\rho (x-a,t)}.
\end{eqnarray}%
The master equation (\ref{Master444}) for the mean density $\rho $ can be
written as
\begin{eqnarray}
\frac{\partial \rho }{\partial t} &=&r(\rho (x-a,t))i(x-a,t)  \notag \\
&&+l(\rho (x+a,t))i(x+a,t)-i(x,t).  \label{ma99}
\end{eqnarray}%
To consider the diffusion approximation when $a\rightarrow 0$ it is
convenient to define the flux of particles $J_{x,x+a}$ from $x$ to $x+a$ as
\begin{equation*}
J_{x,x+a}=r(\rho (x,t))i(x,t)a-l(\rho (x+a,t))i(x+a,t)a
\end{equation*}%
and the flux of particles $J_{x-a,x}$ from $x-a$ to $x$ as%
\begin{equation*}
J_{x-a,x}=r(\rho (x-a,t))i(x-a,t)a-l(\rho (x,t))i(x,t)a
\end{equation*}%
The master equation (\ref{ma99}) can be rewritten in the form%
\begin{equation}
\frac{\partial \rho }{\partial t}=\frac{-J_{x,x+a}+J_{x-a,x}}{a}.  \notag
\end{equation}%
In the limit $a\rightarrow 0$ we obtain
\begin{equation}
\frac{\partial \rho }{\partial t}=-\frac{\partial J(x,t)}{\partial x},
\label{diver}
\end{equation}%
where
\begin{equation}
J(x,t)=-\frac{a^{2}}{2}\frac{\partial i}{\partial x}-\frac{a^{2}i}{1-\rho }%
\frac{\partial \rho }{\partial x}+o(a^{2}).  \label{flux3}
\end{equation}%
\ In the Markovian case, when the escape rate $\gamma (x)$ does not depend
on the residence time and
\begin{equation}
i(x,t)=\gamma (x)\rho (x,t)  \notag
\end{equation}%
we obtain the nonlinear diffusion equation for $\rho $:
\begin{equation}
\frac{\partial \rho }{\partial t}=\frac{\partial }{\partial x}\left[ D\left(
\rho \right) \frac{\partial \rho }{\partial x}\right]  \label{ne}
\end{equation}%
with the nonlinear diffusion coefficient
\begin{equation}
D\left( \rho \right) =\frac{a^{2}\lambda (x)}{2}\frac{1+\rho }{1-\rho }.
\notag
\end{equation}%
In the anomalous subdiffusive case, when the total escape rate from the
point $x$ is given by
\begin{equation*}
i\left( x,t\right) =\frac{1}{\tau _{0}{}^{\mu (x)}}\mathcal{D}_{t}^{1-\mu
(x)}\rho \left( x,t\right) ,
\end{equation*}%
the flux of particles (\ref{flux3}) involves two terms%
\begin{eqnarray}
J(x,t) &=&-\frac{\partial }{\partial x}\left[ D_{\mu }(x)\mathcal{D}%
_{t}^{1-\mu (x)}\rho \left( x,t\right) \right]  \notag \\
&&-\frac{2D_{\mu }(x)\mathcal{D}_{t}^{1-\mu (x)}\rho \left( x,t\right) }{%
1-\rho }\frac{\partial \rho }{\partial x},  \label{flux4}
\end{eqnarray}%
where $D_{\mu }(x)$ is the fractional diffusion coefficient defined by (\ref%
{DD}). Substitution of (\ref{flux4}) into (\ref{diver}) gives a nonlinear
fractional equation
\begin{align}
\frac{\partial \rho }{\partial t}& =\frac{\partial ^{2}}{\partial x^{2}}%
\left[ D_{\mu }(x)\mathcal{D}_{t}^{1-\mu (x)}\rho \right]  \notag \\
& +\frac{\partial }{\partial x}\left[ \frac{2D_{\mu }(x)\mathcal{D}%
_{t}^{1-\mu (x)}\rho \left( x,t\right) }{1-\rho }\frac{\partial \rho }{%
\partial x}\right] .  \label{newf}
\end{align}%
The first term on the RHS of (\ref{newf}) is the standard term for the
subdiffusive fractional equation, while the second one is the nonlinear term
describing the volume filling effect in the subdiffusive case.

\section{Anomalous subdiffusive case with nonlinear tempering}

The aim of this section is to derive the master equation that describes the
transition from subdiffusive transport to asymptotic normal
advection-diffusion transport. We consider the subdiffusive case for which
the waiting time PDF $\psi (x,\tau )$\ has a power law tail: $\psi (x,\tau
)\sim (\tau _{0}/\tau )^{1+\mu (x)}$ with $0<\mu (x)<1$ as $\tau \rightarrow
\infty $. One can use the survival probability $\Psi (x,\tau )$ defined by (%
\ref{sub2}). The advantage of the Mittag-Leffler function (\ref{sub2}) is
that one can obtain the fractional subdiffusive equation without passing to
the long-time limit \cite{Ma}. The Laplace transforms of $\psi (x,\tau
)=-\partial \Psi (x,\tau )/\partial \tau $ and $\Psi (x,\tau )$ are
\begin{equation}
\hat{\psi}\left( x,s\right) =\frac{1}{1+\left( \tau _{0}s\right) ^{\mu (x)}},
\end{equation}%
\begin{equation}
\hat{\Psi}\left( x,s\right) =\frac{\tau _{0}^{\mu (x)}s^{\mu (x)-1}}{%
1+\left( \tau _{0}s\right) ^{\mu (x)}}
\end{equation}%
and, therefore, the Laplace transform of the memory kernel $K(x,t)$ is%
\begin{equation}
\hat{K}\left( x,s\right) =\frac{\hat{\psi}\left( x,s\right) }{\hat{\Psi}%
\left( x,s\right) }=\frac{s^{1-\mu (x)}}{\tau _{0}^{\mu (x)}}.  \label{La}
\end{equation}%
It follows from (\ref{i2}) that instead of (\ref{ano_escape}) we have
\begin{eqnarray}
i\left( x,t\right) &=&\frac{e^{-\Phi (x,t)}}{\tau _{0}{}^{\mu (x)}}\mathcal{D%
}_{t}^{1-\mu (x)}\left[ e^{\Phi (x,t)}\rho \left( x,t\right) \right]  \notag
\\
&&+\alpha (\rho )\rho (x,t),  \label{ano5}
\end{eqnarray}%
where
\begin{equation}
\Phi \left( x,t\right) =\int_{0}^{t}\alpha \left( \rho \left( x,s\right)
\right) ds.  \label{F}
\end{equation}%
This function plays a very important role in what follows. It can be
referred as a nonlinear tempering. From (\ref{Master444}) and (\ref{ano5})
we obtain the subdiffusive master equation for the density $\rho \left(
x,t\right) $ with the nonlinear tempering $\Phi \left( x,t\right) $

\begin{eqnarray}
\frac{\partial \rho }{\partial t} &=&\int_{\mathbb{R}}\frac{e^{-\Phi (x-z,t)}%
}{\tau _{0}{}^{\mu (x-z)}}\mathcal{D}_{t}^{1-\mu (x-z)}\left[ e^{\Phi
(x-z,t)}\rho \left( x-z,t\right) \right]  \notag \\
&&\times w\left( z|\rho (x-z,t)\right) dz  \notag \\
&&-\frac{e^{-\Phi (x,t)}}{\tau _{0}{}^{\mu (x)}}\mathcal{D}_{t}^{1-\mu (x)}%
\left[ e^{\Phi (x,t)}\rho \left( x,t\right) \right]  \notag \\
&&+\int_{\mathbb{R}}\alpha (\rho (x-z,t))\rho (x-z,t)w\left( z|\rho
(x-z,t)\right) dz  \notag \\
&&-\alpha (\rho (x,t))\rho (x,t).  \label{basic2}
\end{eqnarray}%
Using (\ref{basic2}) one can obtain various fractional subdiffusive
nonlinear equations. In particular, for the jump kernel (\ref{ww}) with (\ref%
{www}), in the limit $a\rightarrow 0$ we obtain from (\ref{basic2}) a
nonlinear fractional equation
\begin{align}
\frac{\partial \rho }{\partial t}& =a^{2}\frac{\partial }{\partial x}\left[
\beta \frac{\partial U}{\partial x}\left( \frac{e^{-\Phi }}{\tau _{0}{}^{\mu
(x)}}\mathcal{D}_{t}^{1-\mu (x)}[e^{\Phi }\rho ]+\alpha (\rho )\rho \right) %
\right]  \notag \\
& +\frac{a^{2}}{2}\frac{\partial ^{2}}{\partial x^{2}}\left[ \frac{e^{-\Phi }%
}{\tau _{0}{}^{\mu (x)}}\mathcal{D}_{t}^{1-\mu (x)}[e^{\Phi }\rho ]+\alpha
(\rho )\rho \right]  \notag \\
& +o(a^{2}),  \label{FFPE}
\end{align}%
where $\Phi $ is defined by (\ref{F}).

\subsection{Transition to asymptotic advection-diffusion transport regime
and aggregation phenomenon}

In this subsection we discuss a transition from a subdiffusive transport
regime to an asymptotically normal advection-diffusion transport regime.
This transition is governed by the nonlinear fractional equation (\ref{FFPE}%
). It involves the exponential factor $e^{-\Phi \left( x,t\right) },$ with $%
\Phi \left( x,t\right) =\int_{0}^{t}\alpha \left( \rho \left( x,s\right)
\right) ds,$ that can be considered as a nonlinear tempering. It generalizes
the linear tempering, where the power law waiting time distribution is
truncated by an exponential factor $\exp (-\alpha t)$ \cite{Temp}. We expect
that in the limit $\Phi \left( x,t\right) \rightarrow \infty $ we recover
the stationary advection-diffusion equation or classical Fokker-Planck
equation, while in the intermediate asymptotic regime, when
\begin{equation}
\Phi \left( x,t\right) <<1
\end{equation}%
we have a transient subdiffusive transport. If we take the limit $\Phi
\left( x,t\right) \rightarrow 0$ in (\ref{FFPE}) for a small escape rate $%
\alpha (\rho ),$ nonuniform distribution of the anomalous exponent $\mu (x)$
in (\ref{ff}) leads to the aggregation of particles at the point of minimum
of $\mu (x)$ \cite{Fedot1,Zubarev}. This phenomenon has been observed in
experiments on phagotrophic protists when \textquotedblleft cells become
immobile in attractive patches, which will then eventually trap all
cells\textquotedblright\ \cite{Protist}. In this case the anomalous exponent
$\mu (x)$ dominates and the gradient of the chemotaxis substance (potential
field) $U(x)$ is irrelevant to the partial distribution of the density of
particles $\rho $. However, in general, this aggregation of particles around
one point is just a transient phenomenon in the time interval
\begin{equation}
\tau _{0}<<t<<\frac{1}{\overline{\alpha }}  \label{inter}
\end{equation}%
where $\overline{\alpha }=\frac{1}{t}\int_{0}^{t}\alpha \left( \rho \left(
x,s\right) \right) ds.$ If we consider the long time limit $t\rightarrow
\infty $ such that $\Phi \left( x,t\right) \rightarrow \infty ,$ then we
obtain the stationary advection-diffusion equations (see next two
subsections on aggregation of particles). The solution of these equations
represents ultimate cells aggregation determined by the chemotaxis substance
(potential field) $U(x)$ and the spatial distribution of anomalous exponent $%
\mu (x).$

\subsubsection{Aggregation of particles in the linear case}

Firstly, we assume that the local escape rate $\alpha $ is independent of
time, such that $\Phi \left( x,t\right) =\alpha (x)t.$ The total escape rate
$i\left( x,t\right) $ defined by (\ref{ano5}) takes the form
\begin{eqnarray}
i\left( x,t\right) &=&\frac{e^{-\alpha (x)t}}{\tau _{0}{}^{\mu (x)}}\mathcal{%
D}_{t}^{1-\mu (x)}\left[ e^{\alpha (x)t}\rho \left( x,t\right) \right]
\notag \\
&&+\alpha (x)\rho (x,t),
\end{eqnarray}%
Its Laplace transform $\hat{\imath}\left( x,s\right) =\int_{0}^{\infty
}e^{-st}i(x,\tau )d\tau $ is
\begin{equation}
\hat{\imath}\left( x,s\right) =\left[ \frac{\left( s+\alpha (x)\right)
^{1-\mu (x)}}{\tau _{0}^{\mu (x)}}+\alpha (x)\right] \hat{\rho}(x,s).
\label{Laaa}
\end{equation}%
In the limit $s\rightarrow 0$ ($t\rightarrow \infty )$ one obtains the
stationary escape rate $i_{st}\left( x\right) $ (if it exists) in terms of
the stationary density
\begin{equation}
\rho _{st}\left( x\right) =\lim_{s\rightarrow 0}s\hat{\rho}\left( x,s\right)
.  \notag
\end{equation}%
It follows from (\ref{Laaa}) that the stationary rate $i_{st}\left( x\right)
$ can be written in the Markovian form
\begin{equation}
i_{st}\left( x\right) =\gamma _{\mu }(x)\rho _{st}\left( x\right) ,  \notag
\end{equation}%
where the effective rate of escape $\gamma _{\mu }(x)$ is
\begin{equation}
\gamma _{\mu }(x)=\frac{\alpha (x)}{\left( \tau _{0}\alpha (x)\right) ^{\mu
(x)}}+\alpha (x).  \notag
\end{equation}%
The essential feature of this rate parameter $\gamma _{\mu }(x)$ is that it
depends on the fractal exponent $\mu (x).$ This escape rate, together with (%
\ref{ii}), leads to the stationary advection-diffusion equation
\begin{eqnarray}
&&\frac{\partial }{\partial x}\left[ 2\beta \frac{\partial U}{\partial x}%
D\left( x\right) \rho _{st}(x)\right]  \notag \\
&&+\frac{\partial ^{2}}{\partial x^{2}}\left[ D\left( x\right) \rho _{st}(x)%
\right] =0,  \label{new1}
\end{eqnarray}%
where the diffusion coefficient $D\left( x\right) =a^{2}\gamma _{\mu }(x)/2$
depends on $\mu (x)$ and $\alpha (x):$
\begin{equation}
D\left( x\right) =\frac{a^{2}\left[ \left( \tau _{0}\alpha (x)\right)
^{1-\mu (x)}+\tau _{0}\alpha (x)\right] }{2\tau _{0}}.  \notag
\end{equation}%
When the product $\tau _{0}\alpha $ is small, the term $\left( \tau
_{0}\alpha \right) ^{1-\mu (x)}$ is dominant in the anomalous case $\mu
(x)<1,$ so $D_{\mu }\left( x\right) $ can be approximated as
\begin{equation}
D\left( x\right) =\frac{a^{2}\left( \tau _{0}\alpha (x)\right) ^{1-\mu (x)}}{%
2\tau _{0}},\qquad \tau _{0}\alpha (x)<<1.  \notag
\end{equation}

Let us find the solution $\rho _{st}(x)$ to (\ref{new1}) in the interval $%
[0,L].$ We use the reflective boundary conditions at \ $x=0$ and $x=L$ which
guarantees the conservation of the total population:
\begin{equation}
\int_{0}^{L}\rho (x,t)dx=1.  \notag
\end{equation}%
We introduce the new function
\begin{equation}
p(x)=D\left( x\right) \rho _{st}(x).  \notag
\end{equation}%
It follows from (\ref{new1}) that this function obeys the equation
\begin{equation}
\frac{\partial }{\partial x}\left[ 2\beta \frac{\partial U(x)}{\partial x}%
p(x)\right] +\frac{\partial ^{2}p(x)}{\partial x^{2}}=0
\end{equation}%
with the solution in the form of the Boltzmann distribution%
\begin{equation}
p(x)=N^{-1}\exp \left[ -2\beta U(x)\right] .  \notag
\end{equation}%
Thus, the steady profile is
\begin{equation}
\rho _{st}(x)=N^{-1}D^{-1}\left( x\right) \exp \left[ -2\beta U(x)\right] ,
\label{sss}
\end{equation}%
where $N$ is determined by the normalization condition $N=\int_{0}^{L}D^{-1}%
\left( x\right) \exp \left[ -2\beta U(x)\right] dx.$

To illustrate how the nonhomogeneous anomalous exponent $\mu (x)$ affects
the aggregation pattern we consider the steady profile (\ref{sss}) in the
interval $[0,1]$ for two cases: (a) $U(x)=0$ and (b) $U(x)=mx$ for which the
anomalous exponent $\mu (x)$ has the form
\begin{equation}
\mu (x)=\mu _{0}\exp \left( -kx\right) ,  \notag
\end{equation}%
where $0<\mu _{0}<1$ and $k\geq 0.$ Fig. 1 shows that for the uniform
distribution of chemotaxis substance $U(x)=0$, the particles have the
tendency to aggregate in the region of the small values of $\mu (x)$ (dashed
line). When the gradient of chemotaxis substance $\partial U(x)/\partial x$
forces the cells (particles) to move from the right to the left ($U(x)=5x$
and $\beta =1$), the steady profile is not monotonic (solid line).

\begin{figure}[tbp]
\includegraphics[scale=0.26]{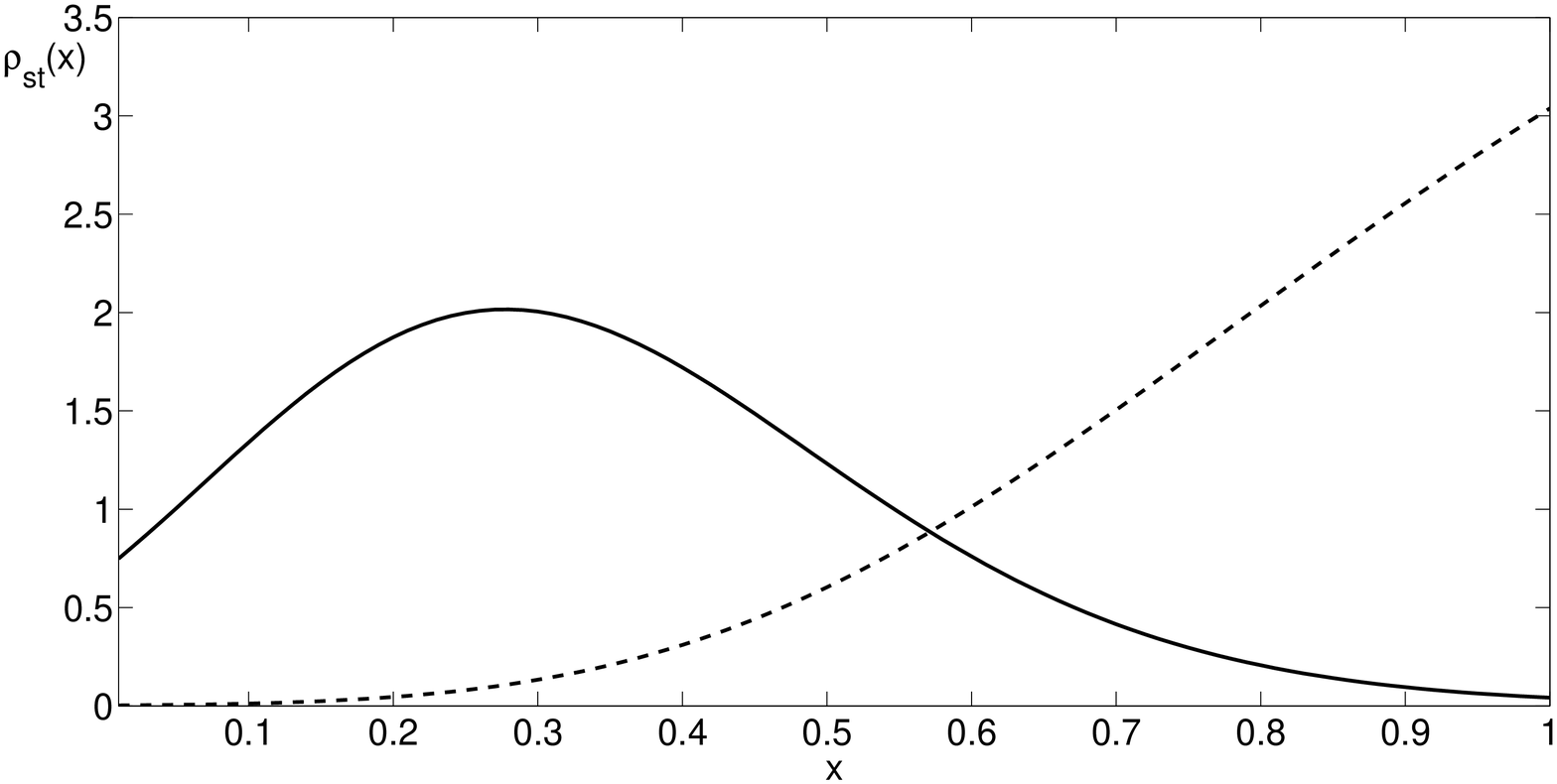}
\caption{{\ Stationary profiles $\protect\rho _{st}(x)$ for the linear
distribution of the potential $U(x)=5x$ (solid line) and $U(x)=0$ (dashed
line), $\protect\beta =1$, $\protect\mu _{0}=0.9$, $\protect\alpha=10^{-4},$
$\protect\tau _{0}=1,$ $k=2.19$. }}
\label{fig1}
\end{figure}

\subsubsection{Aggregation in the nonlinear case}

Now we consider the nonlinear case when the escape rate $\alpha $ depends on
the density $\rho $. For living systems, this nonlinear dependence results
from a coupling between the local density of cells and the intensity of the
response of individual cell to external signals. We assume that in the limit
$t\rightarrow \infty $ the stationary distribution $\rho _{st}(x)$ exists.
Then as $t\rightarrow \infty ,$ the tempering factor $e^{-\Phi \left(
x,t\right) }$ can be approximated by $e^{-\alpha \left( \rho _{st}(x)\right)
t}.$ The stationary escape rate $i_{st}(x)$ corresponding to (\ref{i2}) can
be written in terms of the Laplace transform $\hat{K}\left( x,s\right) $ as
follows
\begin{equation}
i_{st}(x)=\left[ \hat{K}\left( x,\alpha (\rho _{st})\right) +\alpha (\rho
_{st})\right] \rho _{st}\left( x\right) .
\end{equation}%
This steady rate $i_{st}(x)$ has the Markovian form in which the rate
parameter consists of two terms $\hat{K}\left( x,\alpha (\rho _{st})\right) $
and $\alpha (\rho _{st})$. The dependence of the first term on $\alpha (\rho
_{st})$ is due to the non-Markovian character of transport process. This
effect does not exist in the Markovian case for which $\hat{K}$ is a
function of $x$ only. The effective diffusion coefficient $D\left( \rho
_{st}(x)\right) $ is the function of the mean density and it depends on the
structure of the Laplace transform of the memory kernel $K$
\begin{equation}
D\left( \rho _{st}\right) =\frac{a^{2}}{2}\left[ \hat{K}\left( x,\alpha
(\rho _{st})\right) +\alpha (\rho _{st})\right] .  \label{DDD}
\end{equation}

For the subdiffusive case when $\hat{K}\left( x,s\right) $ is defined by (%
\ref{La}), the stationary escape rate $i_{st}(x)$ is%
\begin{equation}
i_{st}(x)=\left[ \frac{\left[ \alpha (\rho _{st}(x))\right] ^{1-\mu (x)}}{%
\tau _{0}^{\mu (x)}}+\alpha (\rho _{st}(x))\right] \rho _{st}\left( x\right)
.  \label{ano}
\end{equation}%
One can see that the first term on the RHS of (\ref{ano}) is dominant for
small $\alpha \tau _{0}$ and $\mu (x)<1$\textbf{. }For the jump density (\ref%
{ww}), in the limit $a\rightarrow 0$ we obtain the stationary nonlinear
Fokker-Planck equation
\begin{equation}
\frac{\partial }{\partial x}\left[ 2\beta \frac{\partial U}{\partial x}%
D\left( \rho _{st}\right) \rho _{st}(x)\right] +\frac{\partial ^{2}}{%
\partial x^{2}}\left[ D\left( \rho _{st}\right) \rho _{st}(x)\right] =0,
\label{st}
\end{equation}%
where $D\left( \rho _{st}(x)\right) $ is the nonlinear diffusion coefficient
defined as
\begin{equation}
D\left( \rho _{st}\right) =\frac{a^{2}\left[ \alpha (\rho _{st}(x))\right]
^{1-\mu (x)}}{2\tau _{0}^{\mu (x)}},\qquad \tau _{0}\alpha (\rho
_{st}(x))<<1.  \notag
\end{equation}%
If we assume a zero flux condition at the boundaries of the interval $[0,L],$
then
\begin{equation*}
J=-2\beta \frac{\partial U}{\partial x}D\left( \rho _{st}\right) \rho
_{st}(x)-\frac{\partial }{\partial x}\left[ D\left( \rho _{st}\right) \rho
_{st}(x)\right] =0,
\end{equation*}%
and the stationary profile $\rho _{st}(x)$ can be found from the nonlinear
equation
\begin{equation}
\rho _{st}(x)=N^{-1}D^{-1}\left( \rho _{st}(x)\right) \exp \left[ -2\beta
U(x)\right] .  \label{st2}
\end{equation}%
where $N$ is determined by the normalization condition $N=\int_{0}^{L}D^{-1}%
\left( \rho _{st}(x)\right) \exp \left[ -2\beta U(x)\right] dx.$ The time
evolution of the density profile $\rho (x,t)$\ for the nonlinear fractional
equation (\ref{FFPE}) can be described as follows. At lower values of $\Phi
=\int_{0}^{t}\alpha (\rho (x,s))ds$, the early evolution is the development
of a single peak at the point of the minimum of $\mu (x).$ This can be
considered as an intermediate anomalous aggregation of particles. However,
incorporating the escape rate $\alpha \left( \rho \right) $ and the
nonlinear tempering factor $e^{-\Phi }$ provide a regularization of
anomalous aggregation. For sufficiently large $\Phi $ the density profile $%
\rho \left( x,t\right) $ must converge to a stationary solution (\ref{st2})
as $t\rightarrow \infty .$

\section{Conclusions.}

The aim in this paper was to derive the macroscopic nonlinear subdiffusive
fractional equations for the evolution of a mean density of random walkers
by incorporating a nonlinear escape rate and nonlinear jump distributions.
The main motivation was to take into account the interaction between
particles on the mesoscopic level at which the random walker characteristics
depend on the mean density of particles. We illustrated the general results
for nonlinear random walk models by using the examples from cell and
population biology. We derived nonlinear fractional equations that take into
account chemotaxis, volume filling effect and cell-to-cell adhesion. We
showed that the nonlinear escape rate leads to the effective regularization
of standard subdiffusive fractional equations. Our modified fractional
equations describe the transition from an intermediate subdiffusive regime
to an asymptotically normal advection-diffusion transport regime. We showed
that this transition is governed by a nonlinear tempering factor that
generalize the standard linear tempering. We discussed the aggregation
phenomenon and showed the impact of a nonuniform distribution of anomalous
exponent on aggregation patterns.

\section{Acknowledgements}

The author gratefully acknowledges the support of the EPSRC Grant
EP/J019526/1 and the warm hospitality of Department of Mathematical Physics,
Ural Federal University. He wishes to thank Steven Falconer and Peter Straka
for interesting discussions.

\end{document}